# Reply to "*Threefold error in the reported zero-field cooled magnetic moment of single crystal La$_2$SmNi$_2$O$_7$* (arXiv: 2602.23240)"

Feiyu Li,[1,15] Zhenfang Xing,[2,15] Di Peng,[2,3*] Jie Dou,[4,5] Ning Guo,[6] Liang Ma,[4,7,8] Yulin Zhang,[1] Lingzhen Wang,[1] Jun Luo,[4] Jie Yang,[4] Jian Zhang,[1] Tieyan Chang,[9] Yu-Sheng Chen,[9] Weizhao Cai,[10,11] Jinguang Cheng,[4,5] Yuzhu Wang,[12] Yuxin Liu,[2] Tao Luo,[2] Naohisa Hirao,[13] Takahiro Matsuoka,[14] Hirokazu Kadobayashi,[13] Zhidan Zeng,[2] Qiang Zheng,[6] Rui Zhou,[4,5] Qiaoshi Zeng,[2,3*] Xutang Tao,[1*] and Junjie Zhang[1*]

[1]State Key Laboratory of Crystal Materials, Institute of Crystal Materials, Shandong University, Jinan, Shandong 250100, China

[2]Center for High Pressure Science and Technology Advanced Research, Shanghai 201203, China

[3]Shanghai Key Laboratory of Material Frontiers Research in Extreme Environments (MFree), Shanghai Advanced Research in Physical Sciences (SHARPS), Shanghai 201203, China

[4]Beijing National Laboratory for Condensed Matter Physics and Institute of Physics, Chinese Academy of Sciences, Beijing 100190, China

[5]School of Physical Sciences, University of Chinese Academy of Sciences, Beijing 100190, China

[6]CAS Key Laboratory of Standardization and Measurement for Nanotechnology, National Center for Nanoscience and Technology, Beijing 100190, China

[7]Key Laboratory of Materials Physics, Ministry of Education, School of Physics, Zhengzhou University, Zhengzhou 450001, China

[8]Institute of Quantum Materials and Physics, Henan Academy of Sciences, Zhengzhou 450046, China

[9]NSF's ChemMatCARS, The University of Chicago, Lemont, Illinois 60439, United States

[10]School of Materials and Energy, University of Electronic Science and Technology of China, Chengdu 611731, Sichuan, China

[11]Huzhou Key Laboratory of Smart and Clean Energy, Yangtze Delta Region Institute (Huzhou), University of Electronic Science and Technology of China, Huzhou 313001, China

[12]Shanghai Synchrotron Radiation Facility, Shanghai Advanced Research Institute, Chinese Academy of Sciences, Shanghai 201204, China

[13]Japan Synchrotron Radiation Research Institute, Sayo, Hyogo 679-5198, Japan

[14]NanoTerasu Promotion Division, Japan Synchrotron Radiation Research Institute (JASRI), Sendai, Miyagi 980-8572, Japan

[15]These authors contributed equally: Feiyu Li and Zhenfang Xing
*email: di.peng@hpstar.ac.cn, zengqs@hpstar.ac.cn, txt@sdu.edu.cn and junjie@sdu.edu.cn

**Abstract**: We respond to the critique by Aleksandr V. Korolev and Evgeny F. Talantsev on the superconducting phase fraction ($f$) calculations in Li *et al*. *Nature* 649, 871-878 (2026). First, the weak upturn in the low-temperature tail of our data has been confirmed to originate from the background, and the paramagnetic Meissner effect is absent in our case; thus, field-cooled (FC) data can be used for superconducting phase fraction calculations. Second, demagnetization effect must be calculated based on the actual measured moment as a function of $f$, which has been well-established and routinely employed in the superconductivity community. In contrast, Korolev and Talantsev treated the demagnetization field as a constant; thus, their calculation significantly underestimates $f$ by a factor of $(1-N\chi_{meas})(1-N)$. This factor is close to 1/3, given $N= 0.849$, $\chi_{meas}= -1.313$ in our study, which explains the origin of their deviated result (nearly three times smaller than our results). Third, our sample is a homogeneous high-quality bulk single crystal, evidenced by various techniques, making the existence of multiple discrete superconducting regions highly unlikely. We conclude that the superconducting phase fraction calculations reported in Li *et al*. *Nature* 649, 871-878 (2026) are not invalidated by the analyses presented in Korolev *et al*. arXiv: 2602.23240 (2026).

We thank Aleksandr V. Korolev and Evgeny F. Talantsev[1] for their interest in our work and for their comments regarding the magnetic characterization of pressurized $La_2SmNi_2O_7$ reported in our *Nature* paper[2]. Specifically, they raised three primary concerns:

*(1). Field-cooled (FC) data cannot be used for superconducting phase fraction calculations due to paramagnetic Meissner effect;*

*(2). They calculated a superconducting phase fraction of 22.8% based on our zero-field-cooled (ZFC) data, almost three times smaller than that reported value of ~62.1% in our Nature paper[2];*

*(3). They proposed that there are an infinite number of combinations of superconducting phase sizes smaller than the physical dimensions of our sample, and questioned the applicability of our method.*

We completely disagree with these assessments and provide our point-by-point response below.

**Response to (1):** The paramagnetic Meissner effect (PME), also known as Wohlleben effect, describes the paramagnetic rather than diamagnetic response in certain superconductors during field-cooling at very low fields[3-8]. However, not all superconductors show PME[5,8]. The defining feature of PME is the observation of positive magnetization upon field-cooling or warming in magnetic susceptibility measurements, which is notably absent in our case[2]. The weak upturn in the low-temperature tail of our data has been confirmed to originate from the background by our experiment[2]. Therefore, there is no valid reason to exclude FC data for calculating the superconducting phase fraction. Actually, FC data have been widely used for superconducting phase fraction calculations, e.g. Ref.[9-11].

**Response to (2):** Korolev and Talantsev[1] calculated the superconducting phase fraction ($f$) using the ratio of the measured total magnetic moment ($m$) to the total ideal Meissner moment of a hypothetical, fully superconducting state ($f = m_{meas}/ m_{ideal, Meissner}$). Their approach implicitly assumes a constant demagnetization field that is independent of the superconducting volume fraction $f$. This assumption is physically inconsistent and leads to a significant underestimation of $f$, as it incorrectly enforces a linear relationship between $f$ and $m_{meas}$ despite the presence of significant demagnetization effects.

In the case of the Meissner state for a superconducting sample with a total volume $V$, measurable total magnetic moment $m$, and constant demagnetizing factor $N$ associated with the applied magnetic field $H_0$ with the actual internal field inside the sample, $H_{int}$ (All in SI units):

$$M = m/V, \quad (1)$$

$$M = \chi H, \quad (2)$$

$$H_{int} = H_0 - NM, \quad (3)$$

By substituting Eq.3 into Eq.2, we obtain:

$$\frac{M}{\chi_{int}} = \frac{M}{\chi_{meas}} - NM, \quad (4)$$

Therefore, we have the equation linking the externally measured susceptibility ($\chi_{meas}$) to the inherent one ($\chi_{int}$):

$$\chi_{meas} = \frac{\chi_{int}}{1 + \chi_{int} N}, \quad (5)$$

In the low-field Meissner state, for a sample with a superconducting shielding fraction $f$, a commonly employed approximation is $\chi_{int} \approx -f$ (neglecting the normal-state contribution post-background subtraction), leading to the conventional self-consistent relations,

$$\chi_{meas} = \frac{-f}{1 - Nf}, \quad (6)$$

$$f = \frac{-\chi_{meas}}{1 - N\chi_{meas}}, \quad (7)$$

Therefore, the measured susceptibility ($\chi_{meas}$) is not linearly proportional to $f$. However, according to Korolev and Talantsev's calculation method[1], $f$ is linear to the $\chi_{meas}$:

$$f = \frac{m_{meas}}{m_{ideal,Meissner}} = \frac{M_{meas}}{\frac{-H_0}{1-N}} = \frac{-M_{meas}}{\frac{H_0}{1-N}} = -\chi_{meas}(1 - N), \quad (8)$$

In the general case, when the shielding fraction $f$ is less than 1 and demagnetization effects are significant, the shielding fraction $f$ must be extracted from the standard self-consistent relation between the measured and intrinsic magnetic responses (Eq.5), which has actually been well-established and routinely employed in the superconductor community[12-22]. Once the demagnetization self-consistency is properly taken into account, one has the correct shielding fraction, consistent with our susceptibility-based analysis[2].

In contrast, based on Eq.8, Korolev and Talantsev's calculation[1] implicitly assumes a constant internal field $H_{int} = H_0/(1-N)$ (namely, a constant demagnetization field), independent of the superconducting volume fraction $f$. This assumption is physically inconsistent and results in a significant underestimation of $f$ by a factor of $(1-N\chi_{meas})(1-N)$. This factor is close to 1/3, given $N$=0.849, $\chi_{meas}$= −1.313 in our study, which explains the origin of their deviated result (nearly three times smaller than our results).

**Response to (3):** We emphasize that the sample employed for our high-pressure magnetic susceptibility measurements is a single crystal with high quality, as verified by energy dispersive spectroscopy, single-crystal X-ray diffraction, nuclear quadrupole resonance, and real-space imaging through scanning transmission electron microscopy (Fig. 2 of our *Nature* paper[2]). The sharp superconducting transition in both resistivity and magnetic susceptibility measurements (Fig. 3 of our *Nature* paper[2]) as well as the uniform electrical transport characteristics (Extended Data Fig. 3d of our *Nature* paper[2]) further attest to the homogeneity of our bulk single crystal sample. Therefore, there is no evidence to support Korolev and Talantsev's proposal that the sample is partitioned into multiple regions of distinct physical dimensions. Furthermore, our shielding fraction estimations among several independent measurements yield consistent values of ~60% (e.g., Fig. 3e, Extended Data Figs. 6a and 6b), which further confirms the homogeneous nature of our high-quality single crystal samples and justifies the use of our method[2].